\documentclass[%
pre,
groupedaddress,
 amsmath,amssymb,
 aps,
hyperref,
twocolumn
]{revtex4-1}

\usepackage{graphicx}
\usepackage{dcolumn}
\usepackage{bm}

\usepackage{color}

\usepackage{multirow}

\begin{document}

\preprint{APS/123-QED}

\title{A Practical Approximation Method for Firing Rate Models of Coupled Neural Networks with Correlated Inputs}

\author{Andrea K. Barreiro}
\affiliation{
 Department of Mathematics\\
Southern Methodist University \\
P.O. Box 750235; Dallas, Texas 75275 U.S.A.
}

\author{Cheng Ly}
\email{CLy@vcu.edu}
\affiliation{%
 Department of Statistical Sciences and Operations Research\\
 Virginia Commonwealth University \\
 1015 Floyd Avenue; Richmond, Virginia 23284 U.S.A.
}%




\date{\today}

\begin{abstract}

Rapid experimental advances now enable simultaneous electrophysiological recording of neural activity at single-cell resolution across large regions of the nervous system.  
Models of this neural network activity will necessarily increase in size and complexity, thus increasing the computational cost of simulating them and the challenge of analyzing them.  
Here we present a novel method to approximate the activity and firing statistics of a general firing rate network model (of Wilson-Cowan type) subject to noisy correlated background inputs.  
The method requires solving a system of transcendental equations and is fast compared to Monte Carlo simulations of coupled stochastic differential equations.  
We implement the method with several examples of coupled neural networks and show that the results are quantitatively accurate even with moderate coupling strengths and an 
appreciable amount of heterogeneity in many parameters.  
This work should be useful for investigating how various neural attributes qualitatively effect the spiking statistics of coupled neural networks.  Matlab code implementing the method is freely available at 
GitHub (\url{http://github.com/chengly70/FiringRateModReduction}).

\end{abstract}


\maketitle


\section{Introduction}

With advances in neural recording technologies, experimentalists can now record simultaneous activity across multiple brain regions at single cell resolution~\cite{ahrens13,prevedel14,brainInit_13,lemon15}. 
However, it is still a technical challenge to measure the interactions within and across brain regions that govern this multi-region activity.  
This challenge is heightened by the fact that cortical neurons are heterogeneous and show substantial trial-to-trial variability~\cite{cohen11}. 
Numerous theoretical studies have examined how neural networks can lead to cortex-like dynamics~\cite{abbottvv,van94,van96,brunel,brunelhakim,buice2007cfa,bressloff09,renart10,touboul11}; 
however, most have been limited to a single region, leaving open the question of how inter-region connection strengths contribute to network processing.  
 
%

One challenge presented by analyzing multi-region neural networks, is the increased number of parameters which must be specified. To survey a high-dimensional parameter space, one must have a way to efficiently simulate (as in \cite{stringer16}) or approximate network statistics (as in \cite{gerstnerBook}). 
Here we present a novel approximation method for calculating the statistics of a general coupled firing rate model (based on~\cite{wilsoncowan1}) of neural networks where we: 
i) assume the \textit{activity} (not the firing statistics) are pairwise normally distributed, 
ii) take the entire probability distribution of the presynaptic neurons/populations (providing input) into account.  Our method is fast because it requires
solving nonlinear equations self-consistently rather than 
simulating 
stochastic differential equations.  Several example neural networks are considered and compared with Monte Carlo simulations.  
A specific version of this method was presented in~\cite{bgsl_17} to model the olfactory sensory pathway; here, we derive formulas in a general way which is easy to 
evaluate and can accommodate heterogeneous networks.  
We also demonstrate the method's efficacy on several example networks with much larger dimension than the specific networks examined in our previous work.

\section{Neural Network Model}  

Each cell (or homogeneous population) has a prescribed activity $x_j$ that is modeled by the following equation~\cite{wilsoncowan1} for $j=1,2,\dots,N_c$:
\begin{eqnarray}\label{xj_eqn}
	\tau_j \frac{dx_j}{dt} = -x_j + \mu_j +\sigma_j \eta_j(t) + \sum_{k=1}^{N_c} g_{jk} F_k(x_k(t))
\end{eqnarray}
where $F_k(\cdot)$ is a transfer function mapping activity to firing rate (in some units), related to the so-called F-I curve, for the $k^{th}$ cell/population.  Thus, the instantaneous firing rate of the $j^{th}$ neuron is:
\begin{eqnarray}\label{frt_eqn}
	F_j(x_j(t)).  
\end{eqnarray}
Depending on the context, the activity variable $x_j$ may represent membrane voltage, calcium concentration, or some other quantity associated with a neuron's internal state \cite{dayan2001theoretical}.
This type of equation has historically been used to capture the average activity of a population of neurons but from here on out we will use the term ``cell" for exposition purposes. 
All cells receive background noise $\eta_j$, the increment of a Weiner process, uncorrelated in time but potentially correlated at each instant: $\langle \eta_j(t) \rangle = 0$, $\langle \eta_j(t) \eta_j(t') \rangle  = \delta(t-t')$, 
and $\langle \eta_j(t) \eta_k(t') \rangle  = c_{jk} \delta(t-t')$ for $j\neq k$ with $c_{jk}\in(-1,1)$.  The parameters $\mu_j$ and $\sigma_j$ are constants that give the background input mean and input standard deviation, respectively.  
The parameter $g_{jk}$ represents coupling strength from the presynaptic $k^{th}$ cell and is a signed quantity; $g_{jk}<0$ represents  inhibitory coupling.

We would like to compute the following statistics: 
\begin{eqnarray}
	\mu(j) &:=& \langle x_j \rangle, \hbox{mean activity}  	\label{stats_defn1} \\
	\sigma^2(j) &:=& \langle x^2_j \rangle - \mu^2(j), \hbox{variance of activity}	\label{stats_defn2} \\
	Cov(j,k) &:=& \langle  x_j x_k \rangle - \mu(j)\mu(k), 		\nonumber \\	
		      &   &		\hbox{covariance of activity} 	\label{stats_defn3} \\ 
	\nu_j &:=& \langle F_j(x_j) \rangle, \hbox{firing rate} \\
	Var(\nu_j) &:=& \langle F^2_j(x_j) -\nu^2_j \rangle, \hbox{variance of spiking} \\
	Cov(\nu_j,\nu_k) &:=& \langle F_j(x_j)F_k(x_k) \rangle - \nu_j\nu_k,	\nonumber \\	
				& &	 \hbox{covariance of spiking} \\
	\rho(\nu_j,\nu_k) &:=& \frac{Cov(\nu_j,\nu_k)}{\sqrt{Var(\nu_j)Var(\nu_k)}},		\nonumber \\
			& &\hbox{correlation of spiking}
\end{eqnarray} 
where the angular brackets $\langle \cdot \rangle$ denote averaging over time and realizations \footnote{We assume the networks of interest satisfy the conditions of the Ergodic Theorems so that averaging over time and state are the same}.  
We will use the following definitions for the following Normal/Gaussian probability density functions (PDF): 
\begin{eqnarray}
	\varrho_1(y) := \frac{1}{\sqrt{2\pi}} e^{-y^2/2},
\end{eqnarray}
the standard normal PDF, and
\begin{eqnarray}
	\varrho_{j,k}(y_1,y_2) := \frac{1}{2\pi\sqrt{1-c_{jk}^2}} \exp\Big( -\frac{1}{2}\vec{y}^T  \left(\begin{smallmatrix} 1 & c_{jk} \\ c_{jk} & 1 \end{smallmatrix}\right)^{-1} \vec{y} \Big), \nonumber \\
\end{eqnarray}
a bivariate normal distribution with $\vec{0}$ mean, unit variance, and covariance $c_{jk}$.  

\textit{In the absence of coupling}, i.e. $g_{jk} = 0$, Eq.~\eqref{xj_eqn} would describe a multi-dimensional Ornstein-Uhlenbeck process. Such a process is well-understood:  any pair of activity variables, $(x_j,x_k)$, are bivariate normal random variables~\cite{gardiner}.  
To see this, consider the following two equations without synaptic coupling:
\begin{eqnarray}
	\tau_j \frac{d x_j}{dt} & = & -x_j + \mu_j + \sigma_j \left( \sqrt{1-c_{jk}}\xi_j(t) + \sqrt{c_{jk}} \xi_c(t) \right) \nonumber \\
			& & \\ 
	\tau_k \frac{d x_k}{dt} & = & -x_k + \mu_k + \sigma_k \left( \sqrt{1-c_{jk}}\xi_k(t) + \sqrt{c_{jk}} \xi_c(t) \right). \nonumber \\	
			& & 
\end{eqnarray}
Note that we have re-written $\eta_{j/k}(t)$ as sums of independent white noise processes $\xi(t)$. 
Since $x_j(t) = \frac{1}{\tau_j}\int_{-\infty}^t e^{-(t-u)/\tau_j} \Big[ \mu_j + \sigma_j\eta_j(u) \Big]\,du$ (where we have taken the initial time to be in the far past to eliminate any impact from the initial conditions), 
we calculate marginal statistics using It\^{o} isometries: 
\begin{equation}
	\mu(j) \equiv \langle x_j  \rangle  =  \mu_j
\end{equation}
\begin{eqnarray*}
	\sigma^2(j) & \equiv & \langle (x_j - \mu(j) )^2 \rangle \\
	 & = & \left\langle   \frac{\sigma^2_j}{\tau_j^2} \int_{-\infty}^t \int_{-\infty}^t e^{-(t-u)/\tau_j} \eta_j(u) e^{-(t-v)/\tau_j} \eta_j(v) \,du\,dv  \right\rangle \\
	 & = &\frac{\sigma^2_j}{\tau_j^2} \int^t_{-\infty} e^{-2(t-u)/\tau_j} \,du=\frac{\sigma^2_j}{2\tau_j}
\end{eqnarray*}

A similar calculation shows that in general we have:
\begin{equation}
	Cov(j,k) = \frac{c_{jk}}{\tau_j+\tau_k} \sigma_j \sigma_k
\end{equation}
Thus, $(x_j,x_k)\sim \mathbb{N}\left( \left(\begin{smallmatrix}\mu_j \\ \mu_k \end{smallmatrix}\right) , \left(\begin{smallmatrix} \frac{\sigma^2_j}{2 \tau_j} & \sigma_j\sigma_k \frac{c_{jk}}{\tau_j+\tau_k} \\ \sigma_j\sigma_k \frac{c_{jk}}{\tau_j+\tau_k} & \frac{\sigma^2_k}{2 \tau_k} \end{smallmatrix}\right) \right)$.  
Statistics for the firing rates, $F(x_j)$, are inherited from this normal distribution, since the firing rate $F(x_j)$ is simply a nonlinear function of the activity $x_j$.

\textit{When coupling is included}, i.e. $g_{jk} \not= 0$ for some indices $j$ and $k$, it may no longer be true that the activity variables $x_j$ remain normally distributed. However, it is reasonable to suppose that, for sufficiently weak coupling, the deviations from a normal distribution will be small. Furthermore, if the firing rate function $F$ has thresholding and saturating behavior (as does a sigmoidal function), then higher moments of $x_j$ have limited impact on statistics of $F(x_j)$. 
Thus, our first assumption will be that each pair of activity variables $(x_j,x_k)$, can be approximated by a bivariate normal, even when coupling is present. We can think of this as a weak coupling assumption, as it holds exactly only with \textit{no} coupling.

\section{Reduction Method}


\begin{table*}
\caption{\label{table:defnsInt} For readability, we define the following quantities.  Whenever $j=k$ in the double integrals (e.g., in $\mathcal{N_F},\mathcal{S}$), the bivariate normal distribution $\varrho_{j,k}$ 
is replaced with the standard normal distribution $\varrho_1$.  Note that order of the arguments matters in $\mathcal{N_F}$: $\mathcal{N_F}(j,k)\neq\mathcal{N_F}(k,j)$ in general; 
all of these quantities depend on the statistics of the activity $\mu(\cdot)$, $\sigma(\cdot)$.}
\begin{ruledtabular}
\begin{tabular}{ll}
Abbreviation & Definition \\ \hline
 $\mathcal{E}_1(k)$ & $\displaystyle \int F_k(\sigma(k)y+\mu(k))\varrho_1(y)\,dy$ \\
 $\mathcal{E}_2(k)$ & $\displaystyle \int F_k^2(\sigma(k)y+\mu(k))\varrho_1(y)\,dy$ \\
$\mathcal{V}(k)$ & $\displaystyle \int F_k^2(\sigma(k)y+\mu(k))\,\varrho_1(y)\,dy - \left(\displaystyle \int F_k(\sigma(k)y+\mu(k))\varrho_1(y)\,dy\right)^2 = \mathcal{E}_2(k)-\left[\mathcal{E}_1(k)\right]^2$ \\
\multirow{ 2}{*}{$\mathcal{N_F}(j,k)$} & $\displaystyle\iint F_k(\sigma(k)y_1+\mu(k))\frac{y_2}{\sqrt{2}}\varrho_{j,k}(y_1,y_2)\,dy_1dy_2, \hbox{if }j\neq k$  \\
& $\displaystyle\int F_j(\sigma(j)y+\mu(j))\frac{y}{\sqrt{2}}\varrho_1(y)\,dy, \hbox{if }j= k$  \\
$\mathcal{S}(j,k)$ & $\displaystyle \iint F_j(\sigma(j)y_1+\mu(j)) F_k(\sigma(k)y_2+\mu(k)) \varrho_{j,k}(y_1,y_2)\,dy_1dy_2$  \\
$\mathcal{C_V}(j,k)$ & $\mathcal{S}(j,k)-\mathcal{E}_1(j)\mathcal{E}_1(k)$  
\end{tabular}
\end{ruledtabular}
\end{table*}

In our method, we assume that time is dimensionless so that the subsequent assumptions have the proper units.  Note that our method can in principle be applied to systems where time has a dimension, as 
long they are of the form in Eq.~\eqref{xj_eqn} (with appropriate units for the parameters).

To compute statistics, we start by writing  Eq.~\eqref{xj_eqn} as a low-pass filter of the right-hand-side: 
\begin{eqnarray}
x_j(t) &= &  x_j (t_0) e^{-(t-t_0)/\tau_j} \nonumber \\
& + &\frac{1}{\tau_j}\int_{t_0}^t e^{-(t-u)/\tau_j} \Big[ \mu_j + \sigma_j\eta_j(u) + \sum_k g_{jk} F_k(x_k(u)) \Big]\,du, \label{eqn:xj_filter} \nonumber \\
\end{eqnarray}
used as the basis for calculating the desired moments of $x_j$.  
For example, when $\langle x_j x_k\rangle$ is desired, we use the previous 
equation for $j$ and $k$, multiply, then take the expected value $\langle \cdot \rangle$. 
By letting the initial time $t_0 \rightarrow -\infty$, we eliminate transients; the resulting statistics will be stationary.  
The resulting exact formulas are 
complicated by the network coupling, so we simplify the calculation(s) as follows.

We only account for direct connections in the formulas for the first and second order statistics, assuming the terms from the indirect connections 
are either small or already accounted for in the direct connections.  For example: although $F_k(x_k(u))$ on the RHS of Eq.~\eqref{eqn:xj_filter} \textit{itself} depends on coupling terms of the form $g_{kl} F_l(x_l)$, etc., we will neglect such terms.
We further make the following assumptions:  
\begin{widetext}
\begin{align}
	& \left\langle \int_{-\infty}^t F_k(x_k(u))e^{-(t-u)/\tau_l}\,du \int_{-\infty}^t F_k(x_k(v))e^{-(t-v)/\tau_{m}}\,dv \right\rangle  \approx  \frac{\tau_l \tau_m}{\tau_l + \tau_m} \mathcal{V}_k + \tau_l \tau_m (\mathcal{E}_1(k))^2   \label{ass_Fvar}  \\
	& \left\langle \int_{-\infty}^t \sigma_j\eta_j(u)e^{-(t-u)/\tau_l}\,du \int_{-\infty}^t F_k(x_k(v))e^{-(t-v)/\tau_{m}}\,dv \right\rangle  \approx  \frac{\tau_l\tau_m}{\tau_l+\tau_m} \sigma_j \mathcal{N_F}(j,k) \label{ass_nzFa} \\
	& \left\langle \int_{-\infty}^t F_j(x_j(u))e^{-(t-u)/\tau_l}\,du \int_{-\infty}^t F_k(x_k(v))e^{-(t-v)/\tau_{m}}\,dv \right\rangle  \approx  \frac{\tau_l \tau_m}{\tau_l + \tau_m} \mathcal{C}_{\mathcal{V}}(j,k) + \tau_l \tau_m \mathcal{E}_1(j)\mathcal{E}_1(k) 		\label{ass_end}
\end{align}
\end{widetext}
See Table~\ref{table:defnsInt} for the definition of the symbols: $\mathcal{E}_2(k), \mathcal{N_F}(j,k), \mathcal{S}(j,k)$.  

Each assumption is equivalent to the assumption that two of the random variables of interest are $\delta$-correlated in time; thus avoiding the need to compute autocorrelation functions explicitly. 
The first assumption, Eq.~\eqref{ass_Fvar}, states that $F_k(x_k(t))$ is $\delta$-correlated with itself; the second, Eq.~\eqref{ass_nzFa}, addresses $\eta_j(t)$ and $F_k(x_k(t))$.
The final assumption, 
Eq.~\eqref{ass_end}, states that $F_j(x_j(t))$  and $F_k(x_k(t))$ are $\delta$-correlated.  
We provide a detailed derivation of Eqs.~\eqref{ass_Fvar}--\eqref{ass_end} in an Appendix (section~\ref{Append_deriv}).  
After testing our method on several examples in section~\ref{sec:examples}, we will revisit the accuracy of these assumptions in 
section~\ref{sec:test_delta}.


%
%
We arrive at the following (approximation) formulas for the statistics of the activity:
\begin{eqnarray} 
	&\mu(j) =  \mu_j + \displaystyle\sum_k g_{jk} \mathcal{E}_1(k) \label{reslt_mn} \\
	& \sigma^2(j) \tau_j =  \frac{\sigma_j^2}{2} +	\sigma_j\displaystyle\sum_k g_{jk} \mathcal{N_F}(j,k) + \frac{1}{2}\displaystyle\sum_k g_{jk}^2 \mathcal{V}(k) \nonumber \\	
	&			 +\displaystyle\sum_{k\neq l} g_{jk} g_{jl} \mathcal{C_V}(k,l)  \label{reslt_vr} \\
	 &Cov(j,k) \frac{\tau_j+\tau_k}{2}  =  \frac{1}{2}c_{jk}\sigma_j\sigma_k + \frac{1}{2}\sigma_j\displaystyle\sum_{l} g_{kl} \mathcal{N_F}(j,l)  \nonumber \\
	 &+ \frac{1}{2}\sigma_k\displaystyle\sum_{l} g_{jl} \mathcal{N_F}(k,l) + \frac{1}{2}\displaystyle\sum_{l_1,l_2} g_{j,l_1} g_{k,l_2} \mathcal{C_V}(l_1,l_2) . \label{reslt_covr}
\end{eqnarray}
See Table~\ref{table:defnsInt} for the definition of the symbols: $\mathcal{E}_1, \mathcal{N_F}, \mathcal{V}, \mathcal{C_V}$, which all depend on the statistical quantities $\mu(\cdot)$ and $\sigma(\cdot)$ of the activity $x_j$.  
Our approximation formulas form a system of 
$\displaystyle\frac{1}{2}\left( N_c^2+3N_c\right)$ equations in $\mu(j)$, $\sigma(j)$, $Cov(j,k)$ (i.e. for the activity only, as defined by Eqs.~\eqref{stats_defn1}--\eqref{stats_defn3}, not the firing) when considering all possible $(j,k)\in\{1,2,\dots,N_c\}$.  
This large system of equations, although nonlinear, is simple to solve because it requires a sequence of function evaluations and matrix multiplications, rather than random sampling. 

Note that the normal distribution assumptions allow us to conveniently write the average quantities as integrals with respect to standard normal distributions but with shifted integrands, which leads to faster calculations because one does not have to 
calculate new probability density functions at each step of the iteration when solving the system self-consistently.  

The resulting formulas can be written compactly with matrices; Eq.~\eqref{reslt_mn} for the mean activity $\mu(j)$ can easily be written as a matrix-vector equation and is thus omitted.  
Let ${\bf Cov}$ denote the $N_c \times N_c$ covariance matrix of the activity 
with ${\bf Cov}(j,k)=Cov(j,k)$, ${\bf G}$ represent the coupling strengths ${\bf G}(j,k)=g_{jk}$, and ${\bf Cr}$ 
denote the correlation matrix of the background noise (i.e. ${\bf Cr}(j,k)=\delta_{jk}+ c_{jk}(1-\delta_{jk})$). Then we have 
\begin{eqnarray}\label{eqnMethod_mat}
	{\bf Cov} =  {\bf IT} \circ \Big( {\bf Cov_0}+{\bf G M_{NF}} + {\bf M_{NF}}^T {\bf G}^T + {\bf G M_{FSq} G}^T \Big) \nonumber \\
\end{eqnarray}
where $\circ$ represents element-wise multiplication, $(\cdot )^T$ denotes matrix transposition, and
\begin{eqnarray}
	& {\bf IT}(j,k) = \frac{1}{\tau_j+\tau_k}  \\
	& {\bf Cov_0}(j,k) = \sigma_j\sigma_k \left[ \delta_{jk} +(1-\delta_{jk}) c_{jk}\right]  \label{eqn:cov_jk}\\
	& {\bf M_{NF} }(j,k) = \sigma_k \mathcal{N_F}(k,j)   \\
	& {\bf M_{FSq} }(j,k) = \mathcal{C_V}(j,k) .
\end{eqnarray}
Note that the matrices ${\bf M_{NF}}$ and ${\bf M_{FSq}}$ have the same nonzero entries as ${\bf Cr}$. 
Denoting ${\bf \Lambda}_{\vec{\sigma}}$ as the diagonal matrix with diagonal $\vec{\sigma}$, the unperturbed covariance (Eq.~\eqref{eqn:cov_jk}) can also be expressed in matrix form as:
\[ {\bf Cov_0} = ({\bf \Lambda}_{\vec{\sigma}}) {\bf Cr} ({\bf \Lambda}_{\vec{\sigma}}) \]

Once the statistics of the activity ($\mu(j)$, $\sigma^2(j)$, and $Cov(j,k)$) are solved for self-consistently, the firing statistics are solved as follows.
\begin{eqnarray}
	\nu_j  &=&  \int F_j( \sigma(j)y+\mu(j))\varrho_1(y)\,dy \label{reslt_frate} \\
	Var(\nu_j) &=& \int F^2_j(\sigma(j)y+\mu(j))\varrho_1(y)\,dy  - \nu^2_j \label{reslt_vRate}
\end{eqnarray}
\begin{widetext}
\begin{eqnarray} \label{reslt_covRate}
	Cov(\nu_j,\nu_k) = \iint F_j(\sigma(j)y_1+\mu(j)) F_k(\sigma(k)y_2+\mu(k)) \mathbb{P}_{j,k}(y_1,y_2)\,dy_1dy_2 - \nu_j \nu_k
\end{eqnarray}
\end{widetext}
where $\mathbb{P}_{j,k}$ is a bivariate normal PDF with zero mean and covariance: 
$\displaystyle\left(\begin{smallmatrix} 1 & \frac{Cov(j,k)}{\sigma(j)\sigma(k)} \\ \frac{Cov(j,k)}{\sigma(j)\sigma(k)} & 1 \end{smallmatrix}\right). $
The off-diagonal terms are obtained from the second order statistics of the activity, Eq.~\eqref{reslt_vr}--\eqref{reslt_covr}.   

\section{Example Networks and Results}\label{sec:examples}

\underline{{\bf Network I.}} We first consider a network that allows us to systematically explore algorithm performance as two key parameters vary.  Specifically, we consider two cells ($N_c=2$) that are reciprocally coupled without autaptic (i.e. self) coupling.  For simplicity, we set the intrinsic parameters for the two cells to be identical, with 
$\tau_j=1$, $F_j(x)=0.5(1+\tanh((x-0.5)/0.1))\in[0,1]$ (arbitrary units), but the mean and variance of the background input differ: $\mu_1=0.15$, $\mu_2=4/15\approx 0.2667$, $\sigma_1=2$, $\sigma_2=3$.  
We vary two parameters: $g_{12}\in[-2,2]$ (input strength from $x_2$ to $x_1$), and $c_{12}=c_{21}\in[0,0.8]$, with $g_{21}=0.4$ fixed.  

\begin{figure*}[tb] 
\begin{center}
	\includegraphics[width=6in]{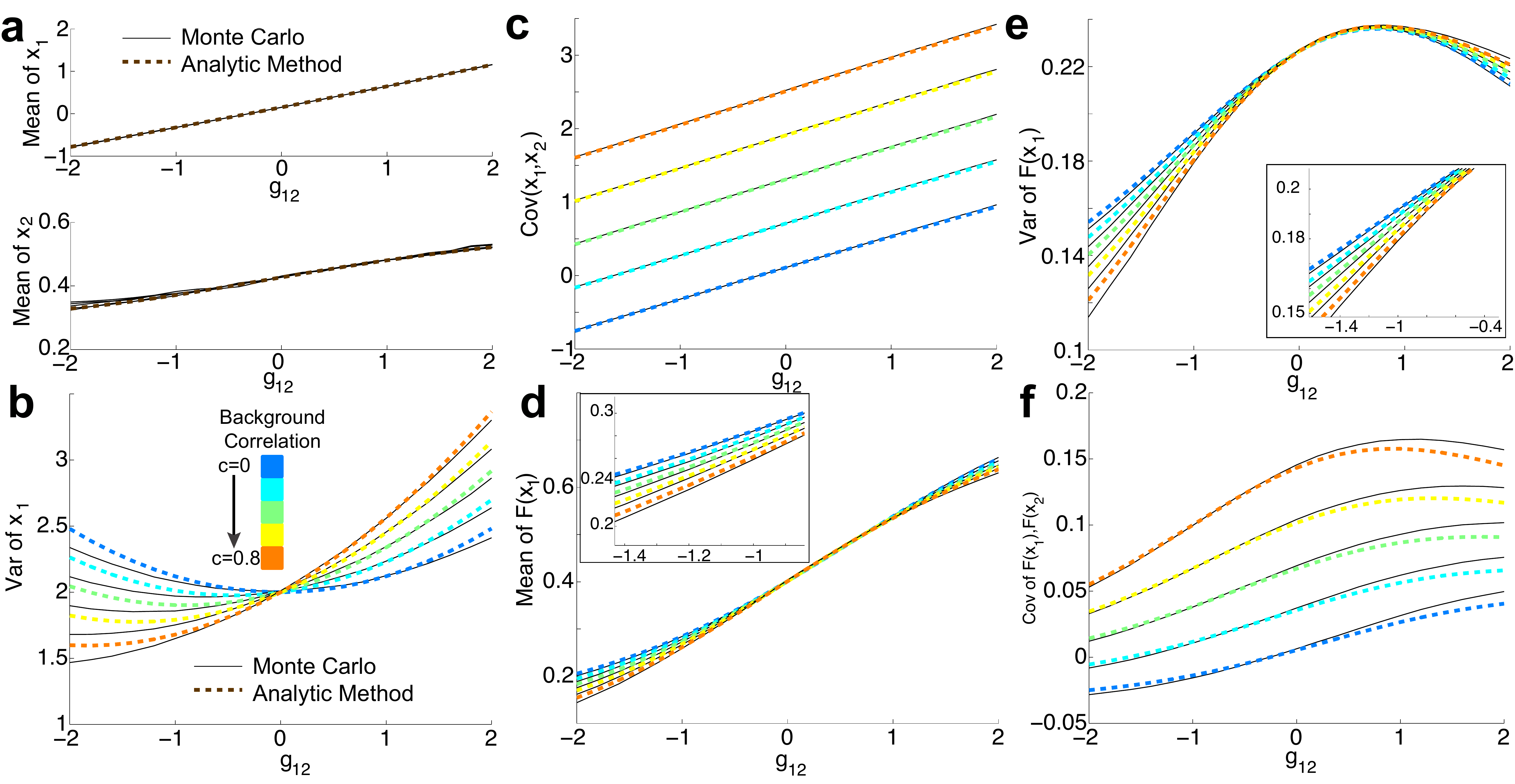}
\end{center}
\caption{Illustration of the method on a network with 2 neurons.  
In all panels, the Monte Carlo simulation results are the thin black solid lines, and the result of the analytic method (Eq.~\eqref{reslt_mn}--\eqref{reslt_covr} solved self-consistently, and 
Eq.~\eqref{reslt_frate}--\eqref{reslt_covRate}) 
are the dashed colored lines representing different background correlation levels.  All parameters are fixed except $g_{12}$ and $c_{12}=c_{21}=:c$; see main text ({\bf Network I}) for values.  
(a) The average activity $x_1$ (top), $x_2$ (bottom) as a function of $g_{12}$ match very well; here the analytic method is in 1 color (brown) because the result is independent of background correlation.  (b) The variance of $x_1$, $\sigma^2(1)$, 
varies with both background correlation and input strength.  The match is very good around $g_{12}=0$ and starts to deviate as $|g_{12}|\to 2$ because with stronger coupling the normal distribution assumption is severely violated.  
(c) The covariance of the activity $Cov(1,2)$.  (d) Mean firing rate: $F(\nu_1)$ slightly depends on $c$; inset is a zoomed-in picture to show that the method captures the relationship of the curves.  (e) Variance of $F(\nu_1)$.  
(f) The covariance of the firing rate $Cov\left(F(\nu_1),F(\nu_2)\right)$.  The corresponding plots for $x_2$ (i.e., panels (b), (d), (e)) are not shown because they do not vary as much, however the analytic method accurately captures the results from Monte Carlo simulations.
\label{fig1}}
\end{figure*}

In Fig.~\ref{fig1}, we see that all of the activity {\it and} firing statistics are accurate compared to Monte Carlo simulations.  Figure~\ref{fig1}(a) shows the mean of $x_1$ as 
the input strength $g_{12}$ varies from negative (inhibitory) to positive (excitatory); this statistic is independent of background correlation.  Figure~\ref{fig1}(b) shows the variance of 
$x_1$; deviations are apparent when the magnitude of the coupling $g_{12}$ is large.  The covariance of the activity (Fig.~\ref{fig1}(c)) is also accurate.  
Even the statistics of the firing rate are relatively accurate; the mean firing rate $F(x_1)$ (Fig.~\ref{fig1}(d)) is only weakly dependent on background correlation whereas the variance of $F(x_1)$ (Fig.~\ref{fig1}(e)) appears to 
vary more with background correlation.  In Fig.~\ref{fig1}(f), the strong dependence of the covariance of the firing rate on background correlation is captured by our method.  For brevity, we omit the corresponding statistics for $x_2$;  the 
method performs equally well there.

\underline{{\bf Network II.}}   We next consider an all-to-all coupled network of $N_c=50$ neurons with heterogeneity in all parameters.  The parameter values were selected from specific distributions and gave rise to 
quenched variability.  The transfer function was set to $F_j(\bullet)=0.5(1+\tanh((\bullet-x_{rev,j})/x_{sp,j}))\in[0,1]$, where $x_{rev,j}$ and $x_{sp,j}$ are fixed parameters that depend on the the $j^{th}$ neuron.  
The distributions of the parameters for this network are:
\begin{eqnarray}
	\tau_j &\sim& \mathbb{N}(1,0.05^2) \label{parms_net2a}\\
	\mu_j &\sim& 2\mathbb{U}-1 \\ 
	\sigma_j &\sim& \mathbb{U}+1\\
	x_{rev,j} &\sim& \mathbb{N}(0,0.1^2) \\
	x_{sp,j} &\sim& 0.35\mathbb{U}+0.05 
\end{eqnarray}
where $\mathbb{U}\in[0,1]$ is a uniform random variable, and $\mathbb{N}$ is normally distributed with the mean and variance as the arguments.  
The covariance matrix ${\bf Cr}$ of the background noise was randomly selected as follows:
\begin{eqnarray}
	{\bf Cr} = ({\bf \Lambda}_{\vec{d_s}}) {\bf A}^T{\bf A} ( {\bf \Lambda}_{\vec{d_s}}) \label{covParms_net2}
\end{eqnarray}
where the entries of the $N_c\times N_c$ matrix ${\bf A}$ are independently chosen from a normal distribution: $a_{j,k}\sim\mathbb{N}(0,0.8^2)$ and $\vec{d_s}$ is the inverse square-root of the diagonal of 
${\bf A}^T{\bf A}$; i.e., if we set ${\bf B}:={\bf A}^T{\bf A}$ with entries $b_{jk}$, then $d_s(j)=1/ \sqrt{b_{jj}}$.  
By construction, ${\bf Cr}$ is symmetric positive semidefinite with 1's on the diagonal.

Finally, the entries of the coupling matrix ${\bf G}$ are randomly chosen, but the parameters of the distribution were varied:
\begin{eqnarray} \label{coupParms_net2}
	{\bf G}(j,k) \sim \mathbb{N}(0,v_l)
\end{eqnarray}
where $v_l=(l/10)^2$ for $l=1,2,3,4$.  There are no zero entries in ${\bf G}$ (i.e. coupling is all-to-all), with both inhibition, excitation, and autaptic (self) coupling.  

\begin{figure*}[tb] 
\begin{center}
	\includegraphics[width=6in]{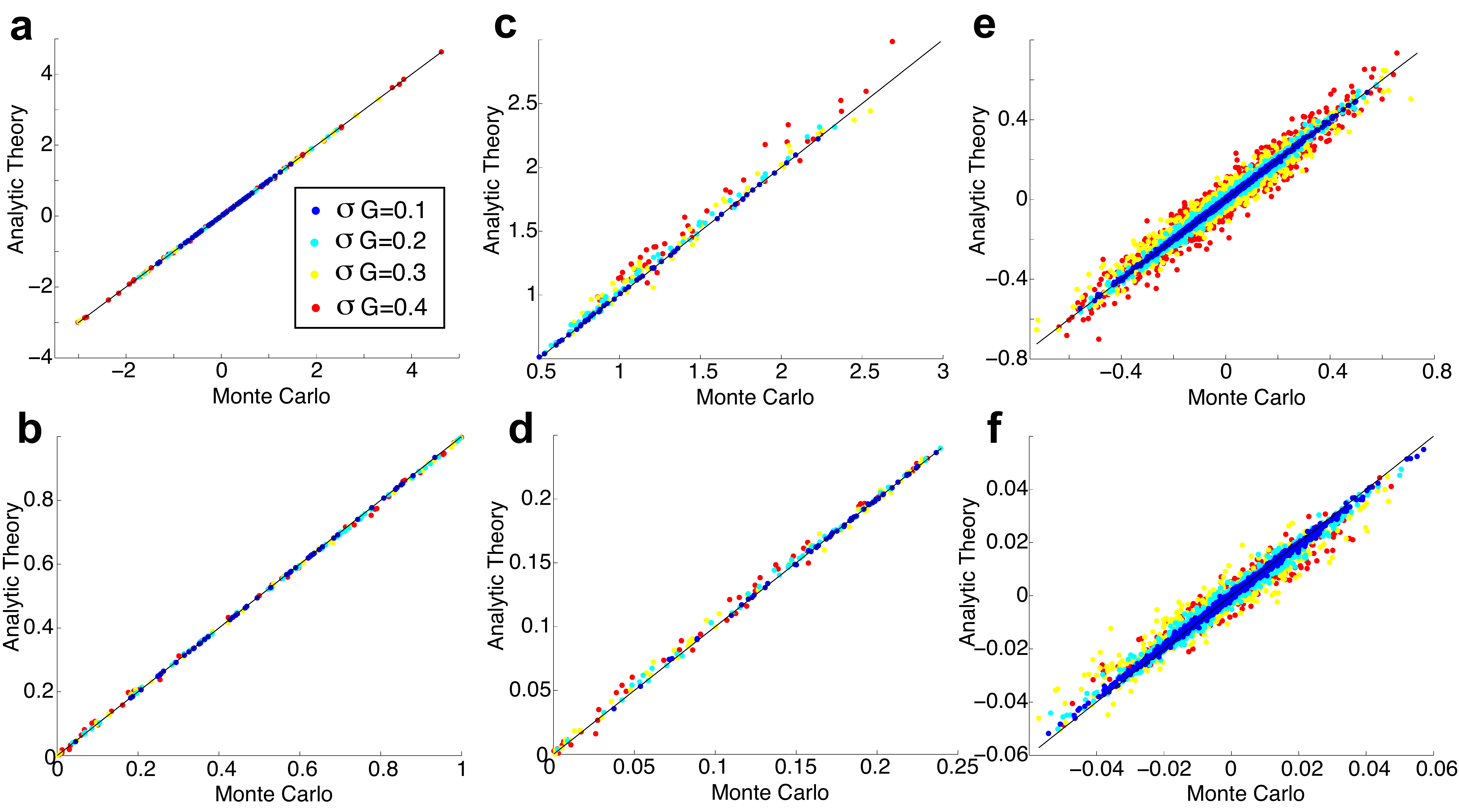}
\end{center}
\caption{A network of $N_c=50$ neurons with heterogeneity in all parameters and all-to-all coupling ({\bf Network II}).  
See Eq.~\eqref{parms_net2a}--\eqref{coupParms_net2} for the distributions of the randomly selected parameters.  
In each panel, four different values of the variance of the distribution of the coupling matrix entries are shown, while the other parameters are held fixed.  
(a) Comparison of the mean activity $\mu(j)$ calculated via Monte Carlo simulations (horizontal axis) and our reduction method (vertical axis), showing all 50 values for each color (coupling matrix distribution).  
(b) Similar to (a) but for mean firing rate $\nu_j$.  
(c) Variance of activity $\sigma^2(j)$.  
(d) Variance of firing rate $Var(\nu_j)$.  
(e) Covariance of activity $Cov(j,k)$, showing all 50*49/2=1225 values for each coupling matrix.  
(f) Covariance of the firing rate $Cov(\nu_j,\nu_k)$.  
The method is accurate but starts to deviate as the overall coupling strength ${\| \bf G \|}$ increases (from blue to red, more deviations from diagonal line).  
\label{fig2}}
\end{figure*}

For each of the four values for the variance of the normal distribution, we chose a single realization of a coupling matrix {\bf G} and 
computed first and second-order statistics of $x_k$ and $F(x_k)$. In Fig.~\ref{fig2} we compared our analytic vs. Monte Carlo results for each cell or cell pair. 
Each realization is identified by a different color; in Fig.~\ref{fig2}(a) for example, there are $N_c$ red data points, corresponding to each $\mu(j)$ for $j=1,...,N_c$. Points that are on the black diagonal line represent a perfect match between Monte Carlo simulations and our method.

First-order statistics $\mu(j)$ and $\nu_j$ are well-captured by the analytic method, even for the largest coupling strength (Fig.~\ref{fig2}(a,b)). This excellent agreement is present despite the substantial amount of heterogeneity in these networks: 
note that $x_j=\mathcal{O}(1)$ and that $F_j\in[0,1]$, and thus that single-cell firing rates in Fig.~\ref{fig2}(b) have a relatively large range.  
Second-order statistics (variances and covariances: Fig.~\ref{fig2}(c-f)) are captured well for smaller coupling values (blue and cyan) but become less accurate for the largest coupling value (red). In particular, the analytic method appears to 
overestimate variance for the largest coupling strength (Fig.~\ref{fig2}(c)).

\begin{figure}[tb] 
\begin{center}
	\includegraphics[width=3.25in]{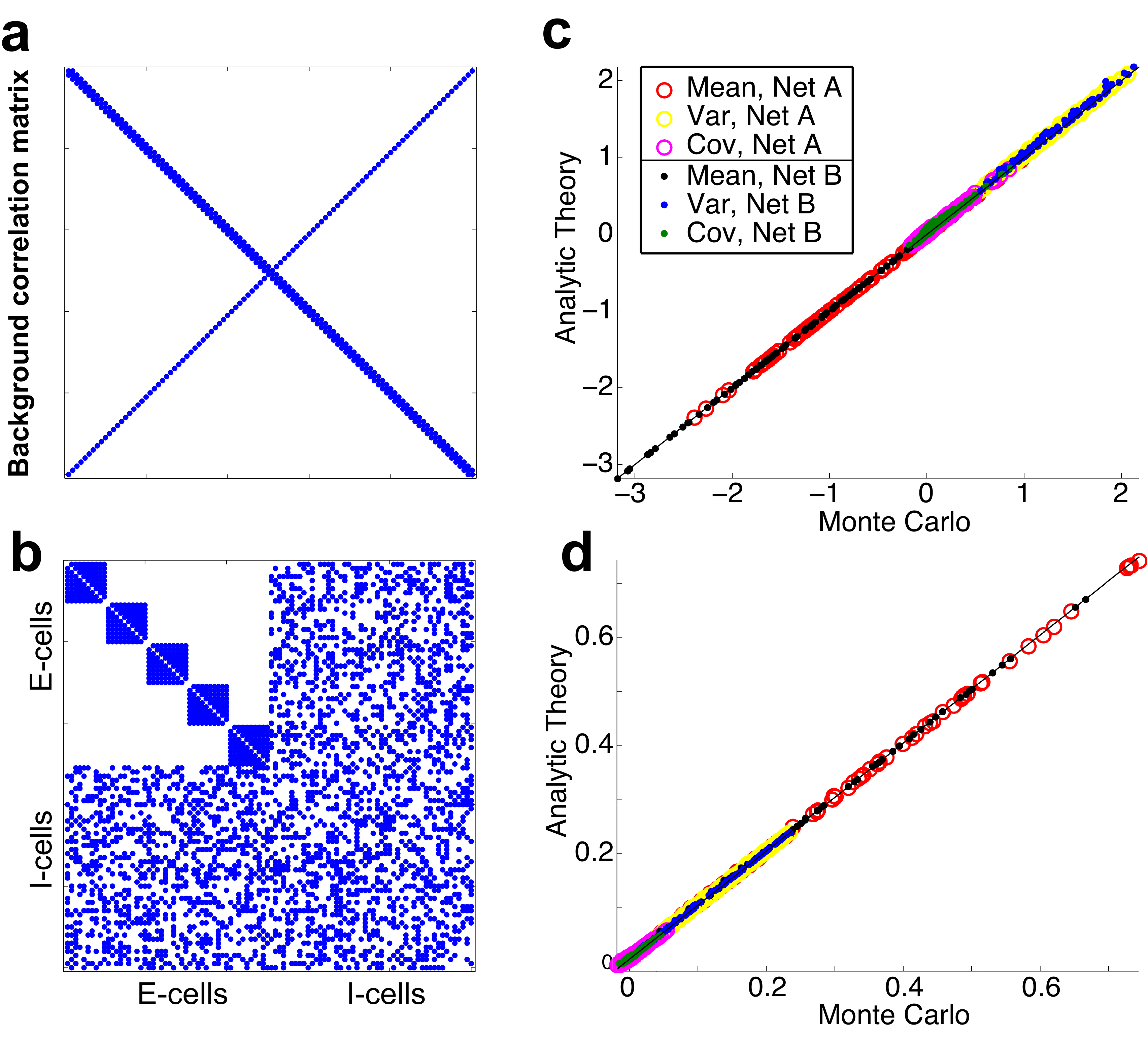}
\end{center}
\caption{A network of $N_c=100$ neurons with heterogeneity in all parameters, but sparse background correlation and clustered and random connectivity ({\bf Network III}).  
See Eq.~\eqref{parms_cr_net3a}--\eqref{parmsInt_net3_end} for the distributions of the randomly selected parameters.  
Sparsity structure of the background correlation matrix ${\bf Cr}$ (a) and coupling matrix ${\bf G}$ (b).
(c) Comparing all of the statistics of the activity for 2 realizations of the network: coupling parameters for network A are: $(g_{EE}, g_{EI},g_{IE},g_{II})=(0.079,-0.24,0.17,-0.31)$ and coupling parameters for network B are: $(g_{EE}, g_{EI},g_{IE},g_{II})=(0.049,-0.38,0.16,-0.17)$.  
As in Fig.~\ref{fig2}, all 100 mean and variance values are plotted, as well as all 4950 covariance values.  
(d) Similar to (c) but for the firing rates.  
\label{fig3}}
\end{figure}

\underline{{\bf Network III.}}  Finally we consider a moderately sized network of $N_c=100$ neurons with quenched heterogeneity in all of the intrinsic parameters, but with more physiological connectivity structure than Network II.  
The first 50 neurons are excitatory $({\bf E})$ ($g_{jk}\geq 0$ for $k=1,2,\dots,50$) and the last 50 are inhibitory $({\bf I})$ ($g_{jk}\leq0$ for $k=51,52,\dots,100$).  
We choose a sparse (random) background correlation matrix via:
\begin{eqnarray}
	c_{jk} = \left\{
\begin{array}{ll}
	1, & \hbox{if }j=k \\
      \mathbb{N}(0.1,0.1^2), & \hbox{if }k=j+1\hbox{ and }j=1,\dots,49 \\
      \mathbb{N}(0.12,0.1^2), & \hbox{if }k=j+1\hbox{ and }j=51,\dots,99 \\
      \mathbb{N}(0.3,0.1^2), & \hbox{if }k=101-j\hbox{ and }j=1,\dots,100 \\
      0 & \hbox{otherwise} \nonumber \\
\end{array} 
\right. 	\\
\label{parms_cr_net3a}
\end{eqnarray}
where as before $\mathbb{N}$ is a Gaussian random variable.  
That is, each cell shares correlated input with its nearest-neighbors of the same type (excitatory vs. inhibitory), and a single cell of the opposite type, where cell location varies along a one-dimensional line. This results in a correlation matrix which is tridiagonal, with an antidiagonal band for the E and I correlation; this sparsity structure is shown in Fig.~\ref{fig3}(a).

In a variety of cortical areas, there is evidence that the correlation of neural activity within a population is on average positive with a wide distribution~\cite{poulet08,yu10,gentet10}; 
thus we set the distributions of excitatory and inhibitory correlation coefficients to $\mathbb{N}(0.1,0.1^2)$ and $\mathbb{N}(0.1,0.12^2)$ respectively (second and third lines of Eq.~\eqref{parms_cr_net3a}).  
Also, there is evidence that E and I neurons are positively correlated (i.e., the synaptic currents are negatively correlated)~\cite{borg98,borg96,okun08}, so we set the average background E-I
correlation ($\mathbb{N}(0.3,0.1^2)$, fourth line of Eq.~\eqref{parms_cr_net3a}) to a higher value than correlations within E or I (second and third lines respectively).  

In order to capture some realistic features of cortical neural networks, we impose sparse but clustered connectivity.  Specifically, we have 5 clusters of E cells of size 10 with all-to-all connectivity and no autaptic (self-coupling) connections, and 
sparse random coupling within the I population (no autaptic connections) and between E and I cells (35\% connection probability).  
See Fig.~\ref{fig3}(b) for the sparsity structure of ${\bf G}$.  This is motivated by experimental evidence that E cells show clustered connectivity~\cite{song05,perin11,ko11}, and
that cells tuned for specific stimulus features can be more connected, while inhibitory connections have less 
structure~\cite{yuste11}.

Synaptic connection strengths were chosen randomly for each realization with the following distributions: 
\begin{eqnarray}
	g_{EE} &=& \mathbb{U}/10,  \nonumber \\
	g_{EI} &=& -\frac{12}{35}\mathbb{U}-\frac{4}{35},  \nonumber \\
	g_{IE} &=& \frac{12}{35}\mathbb{U}+\frac{4}{35},  \nonumber \\		
	g_{II} &=& -\frac{12}{35}\mathbb{U}-\frac{4}{35}, \label{parms_gm_net3}
\end{eqnarray}
where again $\mathbb{U}\in[0,1]$ is a uniform random variable.  
The value $g_{EE}$ is used for all nonzero E to E connections: i.e. $g_{jk}$ with $j,k\in\{1,\dots,50\}$; $g_{EI}$ is used for all nonzero I to E connections: i.e. $g_{jk}$ with $j\in\{1,\dots,50\}$ and $k\in\{51,\dots,100\}$; 
$g_{IE}$ for all nonzero E to I: i.e. $g_{jk}$ with $j\in\{51,\dots,100\}$ and $k\in\{1,\dots,50\}$; similarly for $g_{II}$.

The distributions for the rest of the parameters were similar to {\bf Network II}, with only inconsequential differences:
\begin{eqnarray}
	\tau_j &\sim& \mathbb{N}(1,0.075^2) \\ \label{parmsInt_net3_start}
	\mu_j &\sim& 2\mathbb{U}-1 \\ 
	\sigma_j &\sim& \mathbb{U}+1\\
	x_{rev,j} &\sim& \mathbb{N}(0,0.1^2) \\
	x_{sp,j} &\sim& 0.4\mathbb{U}+0.05  \label{parmsInt_net3_end}
\end{eqnarray}
The choices for $g_{XY}$ and intrinsic parameters are not physiologically motivated, but rather chosen so that we can examine how the algorithm performs on cells with a wide range of intrinsic and network parameters. 

In Fig.~\ref{fig3}(c) and (d) we show the results of the analytic approximation compared to Monte Carlo simulations for the activity and firing rates, respectively.  In each panel, we have combined the mean, variance and covariance and, as in 
Fig.~\ref{fig2}, a data point is plotted for each cell (for means and variances) or cell pair (for covariances). 
Also, we show data from two (2) instances of the network, labeled A and B; for each instance a new realization of the coupling matrix ${\bf G}$ and the coupling parameters (Eq.~\eqref{parms_gm_net3}) are generated 
(see Fig.~\ref{fig3} caption 
for values), but each of the other randomly selected parameters were kept fixed.  
Points that are on the black diagonal line represent a perfect match between Monte Carlo simulations and our method. As with {\bf Network II}, the analytic method accurately captures the statistics cell-by-cell, despite an appreciable degree of 
heterogeneity.

\begin{figure}[htb] 
\begin{center}
	\includegraphics[width=3.25in]{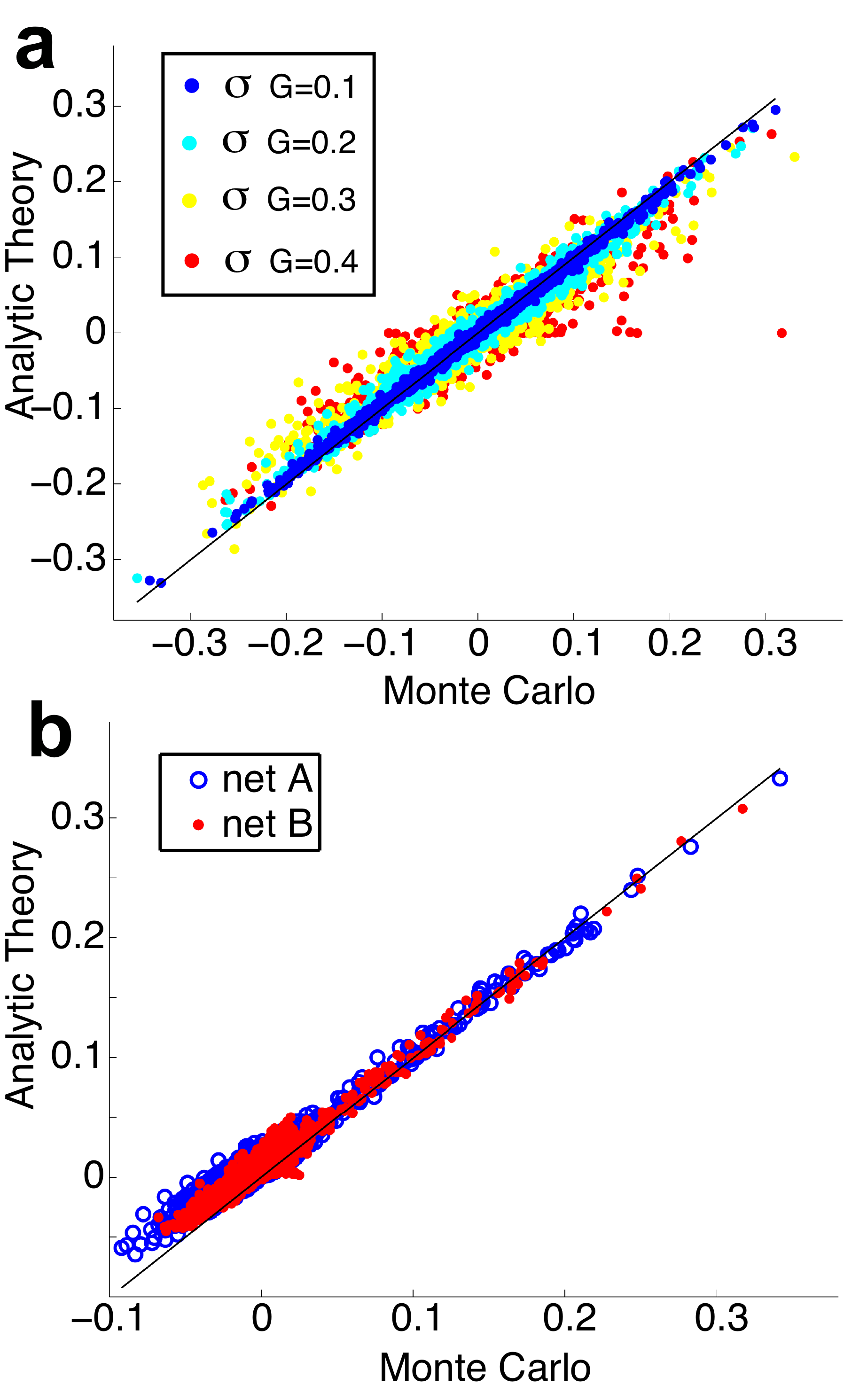}
\end{center}
\caption{Comparisons of the spike count correlation computed by our method and Monte Carlo simulations.  
(a) Comparing the 4 regimes in {\bf Network II}.  The results are accurate because the points predominately lie on the diagonal line.  
As we saw in Fig.~\ref{fig2}, as the relative coupling strength increases, the estimation of the spike count correlation is not as accurate.  
(b) Comparing the 2 networks in {\bf Network III}.  In both cases, the method performs well even though \textit{both} the numerator and denominator are estimated via the method.  
All $N_c(N_c-1)/2$ firing rate correlation values are plotted for each network.  
\label{fig4}}
\end{figure}

Finally, we test how well our method approximates firing rate \textit{correlation}, which is an important normalized measure of trial-to-trial variability (or noise correlations).  The Pearson's correlation coefficient is the predominant measure in 
neuroscience: $\rho(\nu_j,\nu_k)=Cov(\nu_j,\nu_k)/\sqrt{Var(\nu_j,\nu_k)}$, i.e. the ratio of two quantities which we must estimate using the analytic method.  Since this is the ratio of estimated quantities, we might expect larger errors.  
In Fig.~\ref{fig4}, we show comparisons between the analytic method and Monte Carlo simulations for 
{\bf Network II} and 
{\bf Network III}. The method is accurate for a wide range of correlations: Fig.~\ref{fig4}(a) shows correlations as low as $-0.3$ and as high as $0.3$. Thus, the viability of our approximation is not limited to small correlation values, 
but can robustly capture the full range of correlation values observed in cortical neurons~\cite{cohen11,doiron16}.

\section{The $\delta$-correlation Assumption}\label{sec:test_delta}

\begin{figure}[htb!] 
\begin{center}
	\includegraphics[width=3in]{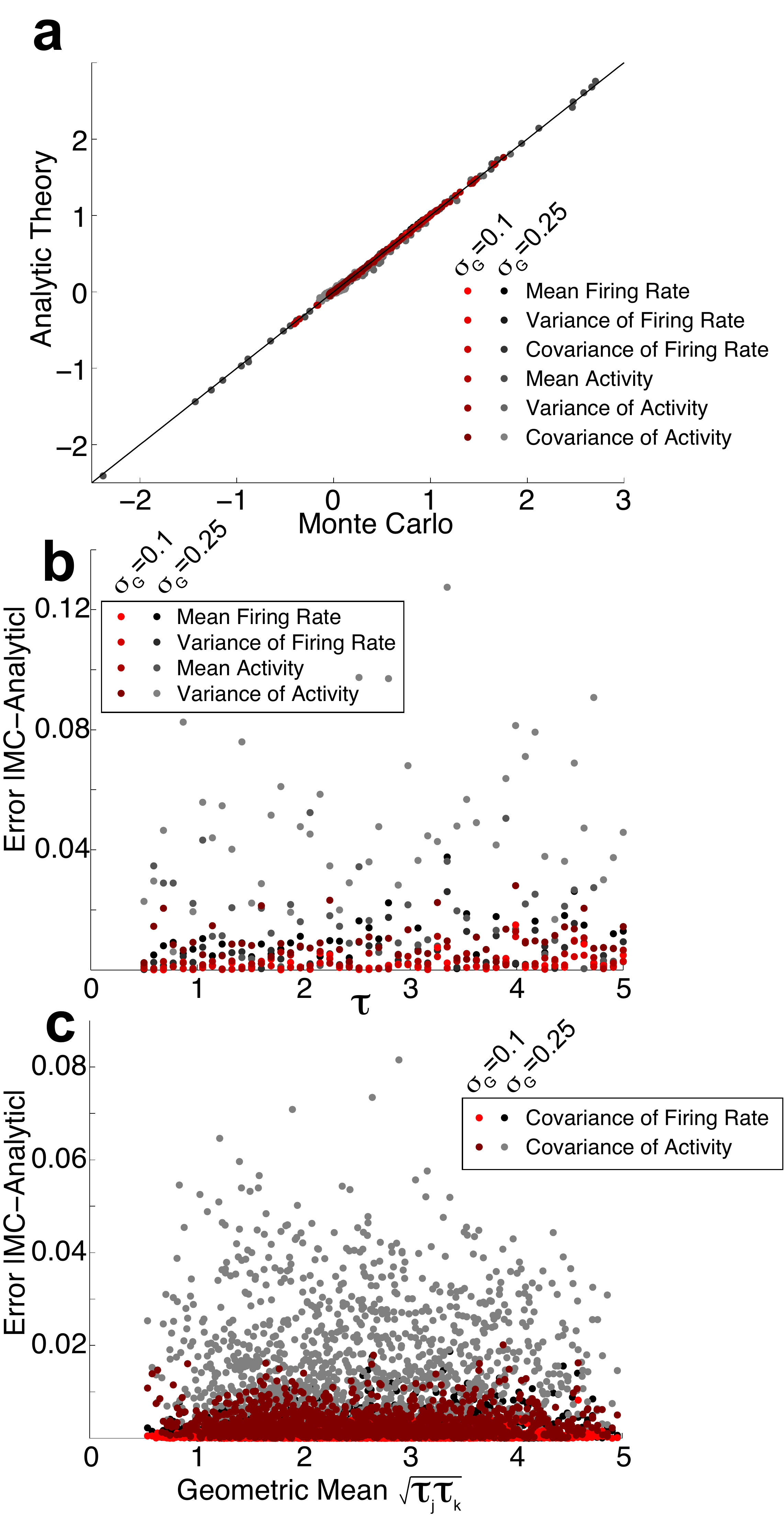}
\end{center}
\caption{A network of $N_c=50$ neurons with heterogeneity predominately in $\tau_j$ (see section {\bf The $\delta$-correlation Assumption} for details).  
(a) Comparing our method to Monte Carlo simulations, plotting all first and second order statistics of the coupled network for two coupling matrices (red shade is with connection strengths ${\bf G}\sim\mathbb{N}(0,0.1^2)$; 
black shade is with ${\bf G}\sim\mathbb{N}(0,0.25^2)$ and is not as accurate).  Despite the large variation in $\tau$, ranging an order of magnitude from 0.5 to 5, the method is accurate.  
(b) Plotting the L$_1$-error of firing rates and variances computed with our method as a function of the $j^{th}$ cell's $\tau_j$ value (see legend).  Note that there is \underline{no trend in error} as 
$\tau_j$ increases.
(c) As in (b), but for all possible covariances of firing rate and activity;  the horizontal axis is the geometric average of the two associated time constants: $\sqrt{\tau_j \tau_k}$.  
The conclusion is the same as in (b), that the $\tau_j$ values are not indicative of the error.
\label{fig5}}
\end{figure}

The assumptions made in deriving Eqs.~\eqref{ass_Fvar}--\eqref{ass_end} --- each equivalent to an assumption that two random variables are $\delta$-correlated in time --- might suggest that the error of our method compared to Monte Carlo simulations would increase as $\tau_j$ increases, or perhaps that the method 
breaks down when the distribution of $\vec{\tau}$ has larger variance.  
Thus far, we have only considered relatively narrow distributions of $\vec{\tau}$.  
We now test this possibility in a setting where we can examine how the method performs as $\tau_j$ is increased without other confounding effects on the error.  

Specifically, we simulate a network of $N_c=50$ cells with the majority of the network parameters set to be 
homogenous values:
$$\mu_j=0.7;\hspace{.05in} \sigma_j=1.3; \hspace{.05in} x_{rev,j}=0.1; \hspace{.05in} x_{sp,j}=0.35,   \forall j $$
The correlation matrix for background noise is a tridiagonal matrix with $0.3$ in the upper and lower diagonal bands, and the coupling matrix is the same as in {\bf Network II}: ${\bf G}(j,k) \sim \mathbb{N}(0,v_l)$ with 
$v_l=(0.1)^2$ and $v_l=(0.25)^2$.  Finally, we set 
\begin{equation}\label{vary_tau_lots}
	\tau_j = (j-1)*\frac{4.5}{N_c-1} + 0.5
\end{equation}
so that $\tau_j$ varies uniformly over an order of magnitude: $\tau_1=0.5$ and $\tau_{N_c}=5$.  

Figure~\ref{fig5}(a) shows that the method is accurate for all possible first and second order statistics despite this large variation in $\vec{\tau}$.  
Figures~\ref{fig5}(b) and (c) show that the L$_1$-error between the Monte Carlo simulations and our method does not depend in any apparent way on the value of the time constant $\tau_j$.  Even when analyzed by a particular statistic, 
there is no trend with time constant.  We conclude that our method is robust to large and disparate values of $\tau_j$.

\section{Accuracy of the reduction method}

\begin{figure}[htb] 
\begin{center}
	\includegraphics[width=3in]{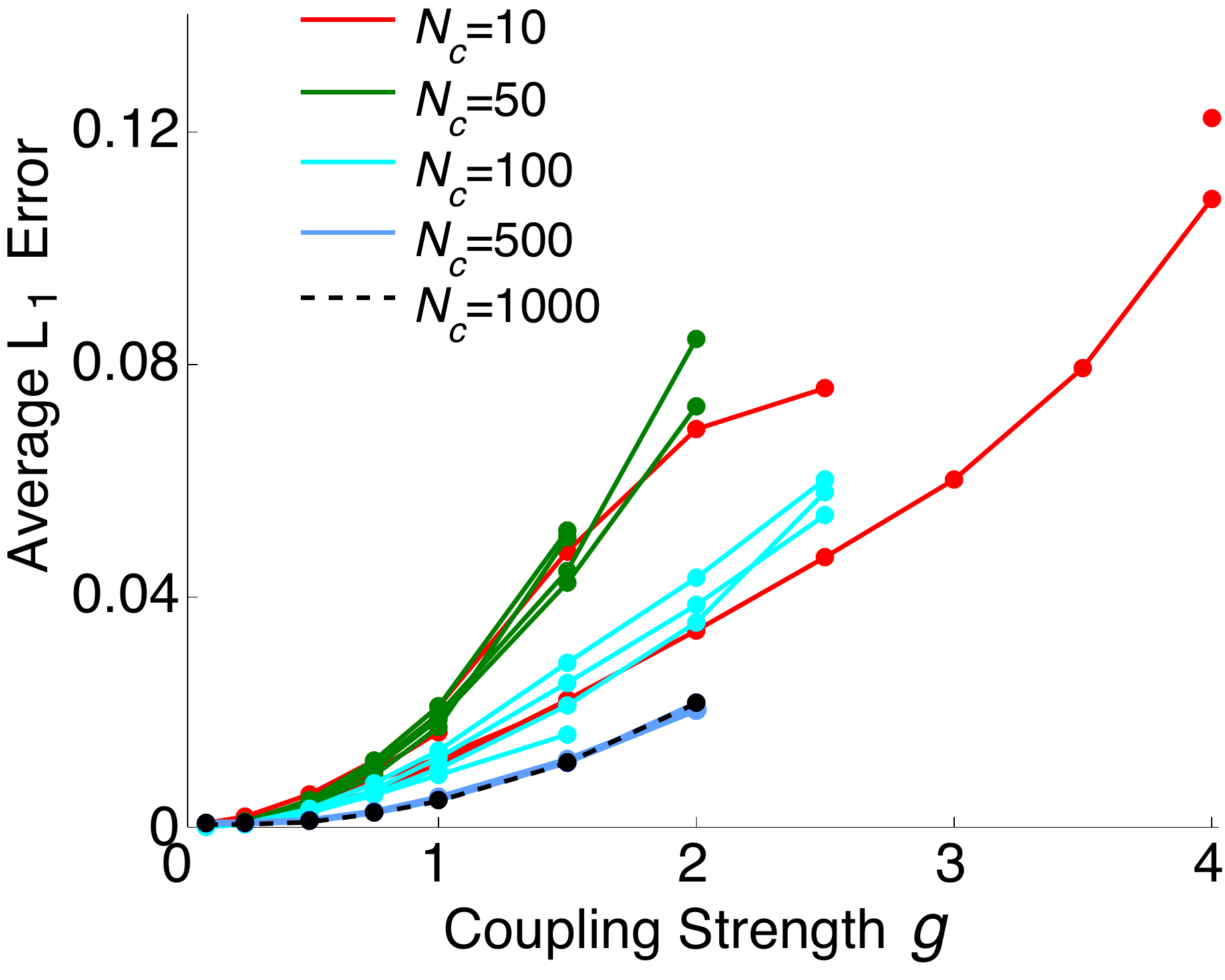}
\end{center}
\caption{Accuracy of the method as coupling strength increases.  
The L$_1$-error (averaged over the entire set of six spiking/activity statistics) of 
our method compared to Monte Carlo simulations increases nonlinearly as coupling strength increases (see main text for definition of $g$ and description of networks), for a variety of network sizes.
As network size $N_c$ increases, the coupling strengths are scaled by $1/\sqrt{N_c}$. 
\label{fig6}}
\end{figure}

As with most calculations that assume small/large values in the parameters, an exact analytic determination of when the approximation fails is difficult, if not impossible. To capture how our method 
deviates as the coupling strength is increased, we performed further computations varying the coupling strengths $g_{jk}$, system size $N_c$, and setting the other parameters to a variety of values.  

The coupling matrix ${\bf G}$ was randomly chosen with half of the entries set to zero, 25\% of the entries set to $g\sqrt{10}/\sqrt{N_c}$ and 25\% set to $-g\sqrt{10}/\sqrt{N_c}$, where $g$ is a scale parameter representing the magnitude of the 
coupling (Note that ${\bf G}$ was chosen only once for each given network size $N_c$, i.e. it was held fixed while other random parameters were varied). The correlation matrix of background noise is a banded matrix with between 1 and 4 bands above/below the diagonal set to $c = 0.3$; that is, each cell shared noise with its $k$-nearest neighbors, for $k=1,2,3$ or $4$.  The rest of the 
parameters were chosen randomly as for {\bf Network III}, Eq.~\eqref{parmsInt_net3_start}--\eqref{parmsInt_net3_end}.

Figure~\ref{fig6}  
shows these results; for each network size $N_c$ and shared common noise footprint $k$, the error of our method compared to Monte Carlo simulations is plotted as a function of the magnitude of the coupling strength $g$.  
Each point represents the L$_1$ error between Monte Carlo and our method, averaged over the entire set of six spiking/activity statistics (mean, variances, and covariances of both activity $x_j$ and firing rate $\nu_j$).  For each system size $N_c$ (except $N_c=1000$; see explanation below), four curves show results for $k=1,2,3$ and $4$ respectively.
We see that as the magnitude of coupling values increase, any given error curve tends to increase.  

In performing computations for Fig. 6, we made the following modifications for computational tractability: for larger $N_c$ values ($N_c=500, 1000$), 
we augmented our method to use a subset of correlation/covariance values; we chose the main diagonal and the super/sub-diagonal $Cov(j,j+1)$; we further computed only one instance for $N_c=1000$ (i.e. there is only one curve). 
We further note that some entries are not plotted because our method did not converge to a solution and/or the resulting covariance matrices are not positive definite, which is not unexpected with randomly chosen parameters.  

In summary, Fig. 6 provides a representative snapshot of the range of possible error values. While the average error increases nonlinearly as 
coupling strength increases, overall the error appears to be relatively insensitive to system size.
We note that the scaling of coupling strengths by the square root of system size $1/\sqrt{N}$ is considered to be ``strong scaling" popularized by the theory of balanced networks \cite{van96,van98}, compared to the relatively weak scaling 
$1/N$.  

\section{Discussion}

There has been a long history of analytic reduction methods for neural network models, both to enhance efficiency in simulation and to aid mathematical analyses.  Here, we summarize some of this literature 
and its relationship to the work presented here.

The simplest approach is a \textit{mean-field} analysis, which would self-consistently estimate the mean values, assuming the variances $\sigma^2(j)$ to be 0 
\cite{tuckwell}. 
However, this neglects the fluctuations which we know to be important in neural systems; therefore many authors have augmented these theories with corrections to capture second-order statistics, higher-order statistics, or time-dependent correlations. 
Several authors have proposed to derive these corrections 
by starting with the microscopic dynamics of single neurons in the network. The microscopic dynamics in question may be given by a master equation~\cite{buice07,bressloff09,buice10,touboul11,bressloff15}, a generalized linear model~\cite{toyoizumi09,ocker2017}, or the theta model~\cite{BC_JSM_2013,BC_PLOSCB_2013}. The result is a principled theory for the second-order statistics of the network, however, the resulting calculations are often complicated and hard to execute.

Here, we aimed to take a middle road between simple (but inaccurate) mean-field calculations, and principled (but complicated) theories to compute network fluctuations from microscopic dynamics.  
Specifically, we start with a system of coupled stochastic differential equations, each of which may represent either a single neuron or a homogeneous population, 
and sought to quickly and accurately estimate statistics of the 
coupled system. Importantly, the unperturbed state in our system is not one in which all neurons are independent; instead we perturb from a state with background noise correlations. Thus, we anticipate this 
approximation can be used to probe a range of neural networks, in which correlations can be significant and activity-modulated.

While the coupled firing rate models we study here were not derived directly from the microscopic dynamics of a spiking network, our results can still yield insight into spiking 
networks~\cite{rosenbaum17,kanashiro17,ledoux11}. 
Our previous work~\cite{bgsl_17} used the qualitative principles and intuitions gained from 
a simple firing rate model to characterize relationships between the analogous parameters in a full spiking model of a multi-region olfactory network.  
In that paper, a small system with simple coupling and background correlations was studied, 
whereas this paper treats networks of arbitrary size, and arbitrary coupling and input correlation structures. The work here is thus a generalization of the calculations in~\cite{bgsl_17}.

In other models, $F(\cdot)$ represents the function that maps firing rate to synaptic input.  Here, we assume that the effective synaptic input $g_{jk}F_k$ is a fixed scaling of the firing rate $F_k$.  
In other biophysical models the effective synaptic input may be a more complex transformation of the firing rate (e.g., an alpha function convolved with firing rate):
the methods presented here can easily be altered to account for this.  
To do this, the only change would be to use $S_k(F_k)$ in Eq.~\eqref{xj_eqn} instead of $F_k$, where $S_k(\cdot)$ is some synaptic activation function.

Our method relies on the assumption that statistics are stationary in time; this assumption allows a set of statistics to be solved self-consistently. 
Thus we have not addressed complex network dynamics, such as oscillations or time-varying statistics.  
However, this limitation is not specific to our method, but also applies to related work. 
Previously developed approximation methods may fail when the system undergoes a bifurcation~\cite{buice07,buice10}, and truncation methods 
(or moment closure methods) are known to fail in certain parameter regimes~\cite{ly_tranchina_07}.  When the set of self-consistent 
equations cannot be solved, there may be other methods available to characterize the oscillatory dynamics (see~\cite{nlc_15} where this is done for the adaptive quadratic integrate-and-fire model).  
Likewise, we did not consider time-lagged network statistics (i.e., the entire cross-correlation functions) but rather only the instantaneous statistics.  This perhaps 
enables the delta-correlation assumption in our method to give accurate approximations even with disparate time-scales (see section~\ref{sec:test_delta}).  
Such considerations are a fruitful path of future work.
%


\begin{acknowledgments}
	Cheng Ly is supported by a grant from the Simons Foundation (\#355173).  
\end{acknowledgments}

%

\section{Appendix: Derivation of Equations (\ref{ass_Fvar})--(\ref{ass_end})}\label{Append_deriv}
Here, we provide a formal derivation of the assumptions that we make to derive our main result Eqs.~\eqref{reslt_mn}--\eqref{reslt_covr}.  Our challenge is that while we are principally interested in zero-time-lag statistics, 
computing second-order statistics (such as $\langle F_j(x_j(t)) F_k(x_k(t)) \rangle$) using Eq.~\eqref{eqn:xj_filter} requires us to know the autocorrelation function, i.e. $\langle F_j(x_j(t)) F_k(x_k(t+\tau)) \rangle$.  
Therefore we need to close the equation by making some kind of assumption about these temporal correlations.

The key assumption to justify Eq.~\eqref{ass_Fvar} is that $F_k(x_k(t))$ is $\delta$-correlated in time: i.e.
\begin{eqnarray}
\left\langle F_k(x_k(u)) F_k(x_k(v)) \right\rangle & = & \delta(u-v) \left(  \left\langle F_k(x_k)^2 \right\rangle - \nu_k^2\right)  + \nu_k^2 \nonumber \\
\end{eqnarray}
where $\nu_k = \left\langle F_k(x_k) \right\rangle$ is the mean.
Using this in the integral on the left-hand side of Eq.~\eqref{ass_Fvar}, we find that
\begin{widetext}
\begin{eqnarray*}
 && \int_{-\infty}^t du \,  \int_{-\infty}^{t} dv\,  e^{-(t-u)/\tau_l} e^{-(t-v)/\tau_m} \,\left\langle F(x_k(u)) F(x_k(v)) \right\rangle  \nonumber\\
& \approx & \int_{-\infty}^t du \,  \int_{-\infty}^{t} dv\,  e^{-(t-u)/\tau_l} e^{-(t-v)/\tau_m} \,  \delta(u-v) \left(  \left\langle F(x_k(u))^2 \right\rangle - \nu_k^2\right) + \int_{-\infty}^t du \,  \int_{-\infty}^{t} dv\, e^{-(t-u)/\tau_l} e^{-(t-v)/\tau_m} \, \nu_k^2 \nonumber\\
  & = & \int_{-\infty}^t du \,   e^{-(t-u)/\tau_l} e^{-(t-u)/\tau_m} \,  \left(  \left\langle F(x_k(u))^2 \right\rangle - \nu_k^2\right) +  \tau_l \tau_m \nu_k^2 \\
  & = & \frac{\tau_l \tau_m}{\tau_l + \tau_m} \left(  \left\langle F(x_k(u))^2 \right\rangle - \nu_k^2\right) + \tau_l \tau_m \nu_k^2 \\
 & = & \frac{\tau_l \tau_m}{\tau_l + \tau_m} \left(  \mathcal{E}_2(k) - (\mathcal{E}_1(k))^2\right) + \tau_l \tau_m (\mathcal{E}_1(k))^2 = \frac{\tau_l \tau_m}{\tau_l + \tau_m} \mathcal{V}_k + \tau_l \tau_m (\mathcal{E}_1(k))^2
\end{eqnarray*}

The other two approximations, Eqs.~\eqref{ass_nzFa} and~\eqref{ass_end}, are arrived at by essentially the same calculation: for completeness, we provide them here as well.

To derive Eq.~\eqref{ass_nzFa}, assume that:
\begin{eqnarray}
\left\langle \sigma_j \eta_j(u) F_k(x_k(v)) \right\rangle & = & \delta(u-v) \left(  \left\langle \sigma_j \eta_j(u) F_k(x_k(u)) \right\rangle -  \langle \sigma_j \eta_j \rangle \langle F_k(x_k) \rangle \right)  + \langle \sigma_j \eta_j \rangle \langle F_k(x_k) \rangle\\
& = & \delta(u-v)  \left\langle \sigma_j \eta_j(u) F_k(x_k(u)) \right\rangle
\end{eqnarray}
where the last line is because $\langle \eta_j \rangle = 0$ and $\sigma_j$ is a constant. Then
\begin{eqnarray*}
\left\langle \int_{-\infty}^t \sigma_j \eta_j(u) e^{-(t-u)/\tau_l} \, du \int_{-\infty}^t F_k(x_k(v)) e^{-(t-v)/\tau_m} \, dv \right\rangle & = &
\int_{-\infty}^t du \,  \int_{-\infty}^{t} dv\,  e^{-(t-u)/\tau_l} e^{-(t-v)/\tau_m} \,\left\langle \sigma_j \eta_j(u) F_k(x_k(v)) \right\rangle \\
& \approx & \int_{-\infty}^t du \,  \int_{-\infty}^{t} dv\,  e^{-(t-u)/\tau_l} e^{-(t-v)/\tau_m} \, \delta(u-v) \left\langle \sigma_j \eta_j(u) F_k(x_k(v)) \right\rangle\\
 & = & \int_{-\infty}^t du \,   e^{-(t-u)/\tau_l} e^{-(t-u)/\tau_m} \, \left\langle \sigma_j \eta_j(u) F_k(x_k(u)) \right\rangle\\
  & = & \frac{\tau_l \tau_m}{\tau_l + \tau_m} \sigma_j \mathcal{N}_F(j,k)
 \end{eqnarray*}
To derive Eq.~\eqref{ass_end}, assume that:
\begin{eqnarray}
\left\langle F_j(x_j(u)) F_k(x_k(v)) \right\rangle & = & \delta(u-v) \left(  \left\langle F_j(x_j(u))F_k(x_k(u)) \right\rangle - \nu_j \nu_k\right)  + \nu_j \nu_k
\end{eqnarray}
where (as before) $\nu_k = \left\langle F_k(x_k) \right\rangle$.
Using this in the integral on the left-hand side of Eq.~\eqref{ass_end}, we find that
\begin{eqnarray*}
& & \left\langle \int_{-\infty}^t F_j(x_j(u)) e^{-(t-u)/\tau_l} \, du \int_{-\infty}^t F_k(x_k(v)) e^{-(t-v)/\tau_m} \, dv \right\rangle\\
& = & \int_{-\infty}^t du \,  \int_{-\infty}^{t} dv\,  e^{-(t-u)/\tau_l} e^{-(t-v)/\tau_m} \,\left\langle F_j(x_j(u)) F_k(x_k(v)) \right\rangle\\
& \approx & \int_{-\infty}^t du \,  \int_{-\infty}^{t} dv\,  e^{-(t-u)/\tau_l} e^{-(t-v)/\tau_m} \,  
\delta(u-v) \left(  \left\langle F_j(x_j(u))F_k(x_k(u)) \right\rangle - \nu_j \nu_k\right)   
  +  \int_{-\infty}^t du \,  \int_{-\infty}^{t} dv\, e^{-(t-u)/\tau_l} e^{-(t-v)/\tau_m} \, \nu_j \nu_k \\
  & = & \int_{-\infty}^t du \,   e^{-(t-u)/\tau_l} e^{-(t-u)/\tau_m} \,  
  \left(  \left\langle F_j(x_j(u))F_k(x_k(u)) \right\rangle - \nu_j \nu_k\right) + \nu_j \nu_k  \tau_l \tau_m\\
  & = & \frac{\tau_l \tau_m}{\tau_l + \tau_m} 
  \left(  \left\langle F_j(x_j(u))F_k(x_k(u)) \right\rangle - \nu_j \nu_k\right)
  + \tau_l \tau_m \nu_j \nu_k \\
 & = & \frac{\tau_l \tau_m}{\tau_l + \tau_m} \left(  \mathcal{S}(j,k) -  \mathcal{E}_1(j)\mathcal{E}_1(k)\right) + \tau_l \tau_m \mathcal{E}_1(j)\mathcal{E}_1(k)\\
 & = & \frac{\tau_l \tau_m}{\tau_l + \tau_m} \mathcal{C}_{\mathcal{V}}(j,k) + \tau_l \tau_m \mathcal{E}_1(j)\mathcal{E}_1(k)
 \end{eqnarray*}
\end{widetext}

\section{Appendix: Alternative Reduction Approaches}

A common method to approximate high dimensional systems is ``moment closure" methods where state variables are integrated or averaged out, and assumptions are made on various 
moments of the random/heterogeneous entities.  Such approaches have a long history in the physical sciences~\cite{chap_cow,dreyer01} and recently in the life 
sciences~\cite{ly_tranchina_07,williams08,buice10}.  Here we provide an alternative approach based on the probability density (or Fokker-Planck) equation of the stochastic neural network, 
rather than the stochastic integrals we considered in the main text. These methods are partially related to the stochastic integral method presented earlier, but we will show that they are {\bf different}.  In a similar vein, we previously showed that 
the analysis of the stochastic integral is more insightful than the Fokker-Planck equation for a system of coupled noisy oscillators (compare main results and Appendix of~\cite{lyErm_pre_10}).

The corresponding probability density function $p(\vec{x},t)$, defined by: $p(\vec{x},t)\,d\vec{x}=P(\vec{X(t)}\in(\vec{x}, \vec{x}+dx))$, of the network models considered in Eq.~\eqref{xj_eqn}
satisfies the following Fokker-Planck equation~\cite{gardiner,risken}:
\begin{widetext}
\begin{eqnarray}
	\frac{\partial p(\vec{x},t)}{\partial t} &=& -\sum_{l=1}^{N_c} \frac{\partial}{\partial x_l}\left\{ \frac{1}{\tau_l}\left[-x_l+\mu_l+\sum_{k=1}^{N_c} g_{lk}F_k(x_k) \right] p(\vec{x},t)\right\}+\frac{1}{2}\sum_{j,k} D_{j,k}\frac{\partial^2 p(\vec{x},t)}{\partial x_j\partial x_k} 
\end{eqnarray}
\end{widetext}
where $D_{j,k}=c_{jk}\frac{\sigma_j\sigma_k}{\tau_j\tau_k}$ and the second sum is taken over all $N_c\times N_c$ pairs of $(j,k)$.  This high-dimensional partial differential equation contains all of the statistics about $\vec{X}$ and any desired 
transformations.  We are interested in the steady-state equation $\frac{\partial p(\vec{x},t)}{\partial t}=0$, assuming the statistics are in equilibrium.  
It is convenient to write the function in the curly brackets as a probability flux or current, as follows:
\begin{equation}\label{flux_defn}
 J_l(\vec{x},t) :=\frac{1}{\tau_l}\left[-x_l+\mu_l+\sum_{k=1}^{N_c} g_{lk}F_k(x_k) \right] p(\vec{x},t).
\end{equation}
The steady-state equation is:
\begin{widetext}
\begin{eqnarray}
0 &=& -\sum_{l=1}^{N_c} \frac{\partial}{\partial x_l}\left\{ \frac{1}{\tau_l}\left[-x_l+\mu_l+\sum_{k=1}^{N_c} g_{lk}F_k(x_k) \right] p(\vec{x})\right\}+\frac{1}{2}\sum_{j,k} D_{j,k}\frac{\partial^2 p(\vec{x})}{\partial x_j\partial x_k}  \\
0 &=& -\sum_{l=1}^{N_c} \frac{\partial}{\partial x_l} J_l(\vec{x})+\frac{1}{2}\sum_{j,k} D_{j,k}\frac{\partial^2 p(\vec{x})}{\partial x_j\partial x_k}  \label{fp_sseqn}
\end{eqnarray}
\end{widetext}

\subsection{Moment Closure Methods}

We want to reduce this high-dimensional system into one that is solvable.  Without coupling $g_{jk}=0$, the solution is simply a multivariate Gaussian distribution with  mean $\vec{\mu}$ and covariance matrix 
$Cov(j,k)=\frac{c_{jk}}{\tau_j+\tau_k}\sigma_j\sigma_k$.  This motivates a closure of the system where we assume $\vec{X}$ is determined by its first two moments and is approximated by a Gaussian: $X_j=\sigma(j)+Y\mu(j)$, 
where $Y$ is a standard normal random variable.  We also assume the joint marginal distributions are bivariate Gaussians:
\begin{equation}\label{margGauss_ass}
	\mathcal{P}(x_j,x_k):=\int p(\vec{x}) \, d\vec{x}_{\backslash j,k}\sim \mathbb{N}_2
\end{equation}
where $\mathbb{N}_2$ is the following bivariate Gaussian distribution: $\mathbb{N}\left( \left(\begin{smallmatrix}\mu(j) \\ \mu(k) \end{smallmatrix}\right) , \left(\begin{smallmatrix} \frac{\sigma(j)^2}{2 \tau_j} & \sigma(j)\sigma(k) \frac{c_{jk}}{\tau_j+\tau_k} \\ \sigma(j)\sigma(k) \frac{c_{jk}}{\tau_j+\tau_k} & \frac{\sigma(k)^2}{2 \tau_k} \end{smallmatrix}\right) \right)$, and 
$d\vec{x}_{\backslash j,k}$ denotes integrating over all $N_c$ variables except $x_j$ and $x_k$.  Note that these assumptions are also made in the main text. 

We multiply Eq.~\eqref{fp_sseqn} by $x_j$ and integrate the equation over all $N_c$ variables, $d\vec{x}=dx_j\,d\tilde{x}$ 
(where $d\tilde{x}=dx_1\dots dx_{j-1} dx_{j+1}\dots dx_{N_c}:=d\vec{x}_{\backslash j}$):
\begin{eqnarray}
	0 &=& - \int \sum_{l=1}^{N_c} \frac{\partial}{\partial x_l} J_l(\vec{x}) x_j \,dx_j\,d\tilde{x} \nonumber \\
	 & &+ \frac{1}{2} \int \sum_{l_1,l_2}D_{l_1,l_2} \frac{\partial^2 p(\vec{x})}{\partial x_{l_1}\partial x_{l_2}} x_j\,dx_j\,d\tilde{x} \label{eqn:first_pde}
\end{eqnarray}
Consider the first term: when $l\neq j$, we have: 
\begin{eqnarray}
\int \frac{\partial}{\partial x_l} J_l(\vec{x}) x_j \,dx_j\,d\tilde{x}&=& \int \frac{\partial}{\partial x_l} J_l(\vec{x}) dx_l\, x_j \,dx_j \,d\vec{x}_{\backslash l, j} \nonumber \\		
					& =&\int J_l\vert_{x_l=-\infty}^{x_l=\infty} x_j\,dx_j \,d\vec{x}_{\backslash j} \nonumber \\	
					& =& \int 0 \, x_j\,dx_j\,d\vec{x}_{\backslash j} = 0
\end{eqnarray}
The last equality comes from no flux at $\pm\infty$: $J_l \vert_{x_l=-\infty}^{x_l=\infty}=0$.  A similar calculation applies to the second term, for all $N_c\times N_c$ values of $(l_1,l_2)$, it is 0.  
When $l_1\neq j$ and $l_2\neq j$, integrate in $x_{l_1}$ and $x_{l_2}$ first and use the fact that there is no density at $\pm\infty$: $p(\vec{x})\vert_{x_{l_{1/2}}=-\infty}^{x_{l_{1/2}}=\infty}=0$; 
when $l_{1/2}=j$, integrate in $x_j$ first then integrate by parts, using  $\partial_j p(\vec{x}) x_j\vert_{x_{l_{1/2}}=-\infty}^{x_{l_{1/2}}=\infty}=0$ and 
$\partial_j p(\vec{x}) \vert_{x_{l_{1/2}}=-\infty}^{x_{l_{1/2}}=\infty}=0$.  Therefore, Eq.~\eqref{eqn:first_pde} 
is:
\begin{eqnarray}
	0 &=& - \int \frac{\partial}{\partial x_j} J_j(\vec{x}) x_j \,dx_j\,d\tilde{x}  \nonumber \\
	0 &=& - \int J_j(\vec{x})x_j\vert_{x_j=-\infty}^{x_j=\infty}\,d\tilde{x} + \int J_j(\vec{x})\,d\vec{x} \nonumber \\	
	0 &=& -0 + \frac{1}{\tau_j}\left( -\mu(j) + \mu_j + \sum_{k=1}^{N_c} g_{jk} \mathcal{E}_1(k) \right) 	\label{eqn:first_pdeEval}
\end{eqnarray}
where $\mu(j) := \int x_j p(\vec{x})\,d\vec{x}$, and we have used the approximation $\int F_k(x_k)p(\vec{x})\,d\vec{x}\approx\mathcal{E}_1(k)$ 
(see Table~\ref{table:defnsInt} for definition of $\mathcal{E}_1(k)$) by assuming the marginal $x_k$ PDF is a normal distribution with mean $\mu(k)$ and variance $\sigma^2(k)$.  Re-arranging 
Eq.~\eqref{eqn:first_pdeEval} gives the exact same nonlinear equation for the mean $\mu(j)$, but coupled with the variance via $\mathcal{E}_1(k)$:
\begin{equation}\label{eqn:meanFP}
	\mu(j) = \mu_j + \sum_{k=1}^{N_c} g_{jk} \mathcal{E}_1(k)
\end{equation}

To derive a similar equation for the variance $\sigma^2(j)$, we multiply Eq.~\eqref{fp_sseqn} by $x_j^2$ and again integrate over all variables:
\begin{eqnarray}
	0 &=& - \int \sum_{l=1}^{N_c} \frac{\partial}{\partial x_l} J_l(\vec{x}) x_j^2 \,dx_j\,d\tilde{x}  \nonumber \\
	    & &+ \frac{1}{2} \int \sum_{l_1,l_2}D_{l_1,l_2} \frac{\partial^2 p(\vec{x})}{\partial x_{l_1}\partial x_{l_2}} x_j^2\,dx_j\,d\tilde{x} \label{eqn:second_pde}
\end{eqnarray}

First consider the diffusion (second) term: similar to before, if either $l_1\neq j$ or $l_2\neq j$, the term will vanish (integrate in $x_{l_1}$ and $x_{l_2}$ first and use the fact that there is no density 
at $\pm\infty$: $p(\vec{x})\vert_{x_{l_{1/2}}=-\infty}^{x_{l_{1/2}}=\infty}=0$).  However, when $l_1=l_2=j$, integrate $x_j$ first and use integration by parts twice to get:
$$ \frac{D_{j,j}}{2}\int \frac{\partial^2 p(\vec{x})}{\partial x_j^2} x_j^2\,dx_j =  D_{j,j} \int p(\vec{x})\,dx_j;$$
taking into account the other $N_c-1$ integration variables and that $\int p(\vec{x})\,d\vec{x}=1$ simply gives $D_{j,j}$ for the second term.

Now for the first term in Eq.~\eqref{eqn:second_pde}: if $l\neq j$, then we can see that that term in the sum vanishes by integrating $x_l$ first and using $J_l \vert_{x_l=-\infty}^{x_l=\infty}=0$.  Using 
integration by parts for the $l=j$ term, we get:
\begin{eqnarray}
&- \int  \frac{\partial}{\partial x_j} J_j(\vec{x}) x_j^2 \,dx_j\,d\tilde{x}=0+2\int x_j J_j(\vec{x})\,dx_j\,d\tilde{x} \nonumber \\
&=\frac{2}{\tau_j} \displaystyle\int \left[ -x^2_j +\mu_j x_j + \displaystyle\sum_{k=1}^{N_c} g_{jk} x_j F_k(x_k) \right]p(\vec{x}) \,dx_j\,d\tilde{x} \nonumber
\end{eqnarray}
Using the fact that $\int x^2_j p(\vec{x})\,d\vec{x}=\sigma^2(j)+\mu(j)^2$, the entire Eq.~\eqref{eqn:second_pde} is:
\begin{widetext}
\begin{equation}
D_{j,j} = \frac{2}{\tau_j}\left[ \sigma^2(j)+\mu(j)^2 - \mu_j \mu(j)-\displaystyle\sum_{k=1}^{N_c} g_{jk}\int x_j F_k(x_k) p(\vec{x})\,d\vec{x} \right] \label{eqn:sec_interm}
\end{equation}
\end{widetext}
This equation is exact thus far.  We now employ our approximation: $X_j=\mu(j)+Y_1\sigma(j)$ where $Y_1$ is a standard normal random variable (similarly for $X_k$).  The last term in the previous equation is:
\begin{eqnarray}
& \mu(j)\displaystyle\sum_{k=1}^{N_c} g_{jk}\int F_k(x_k)p(\vec{x})\,d\vec{x} \nonumber \\
& +\sigma(j)\displaystyle\sum_{k=1}^{N_c} g_{jk}\int y_1 F_k(\mu(k)+y_2\sigma(k))p(\vec{x})\,d\vec{x} \nonumber
\end{eqnarray}
We can approximate the first term above with Eq.~\eqref{eqn:meanFP} to get (excluding $\mu(j)$):
$$\displaystyle\sum_{k=1}^{N_c} g_{jk}\int F_k(x_k)p(\vec{x})\,d\vec{x} \approx \sum_{k=1}^{N_c} g_{jk}\mathcal{E}_1(k) \approx \mu(j)-\mu_j$$
Thus, this leads to a cancellation of the terms $\mu(j)^2$ and $\mu_j\mu(j)$ in Eq.~\eqref{eqn:sec_interm}.  We approximate the term $\int y_1 F_k(\mu(k)+y_2\sigma(k))p(\vec{x})\,d\vec{x}$ by assuming the 
joint marginal distribution of $(X_j,X_k)$ are bivariate normal, and use the definition of $\mathcal{N_F}$ in Table~\ref{table:defnsInt} to get:
$$\int y_1 F_k(\mu(k)+y_2\sigma(k))p(\vec{x})\,d\vec{x} \approx \sqrt{2} \mathcal{N_F}(j,k).$$
Therefore, the equation for the variance is:
\begin{eqnarray}\label{eqn:varFP}
	\sigma^2(j) \tau_j= \frac{\sigma^2_j}{2} +\sigma(j) \sqrt{2}\tau_j \displaystyle\sum_{k=1}^{N_c} g_{jk} \mathcal{N_F}(j,k)
\end{eqnarray}
This equation is similar to Eq.~\eqref{reslt_vr} but lacking higher order terms in $F_k$, as well as other differences.  

To derive the analogous equation for the $Cov(j,k)$, the procedure is almost exactly the same except Eq.~\eqref{fp_sseqn} is multiplied by $x_j x_k$, and there are two terms from the sum (over probability fluxes $J_l$) that contribute, when 
$l=j$ and $l=k$.  The result is:
\begin{eqnarray}
Cov(j,k)\frac{\tau_j+\tau_k}{2} &=& c_{jk}\frac{\sigma_j\sigma_k}{2} +\frac{\sigma(j)}{2} \sqrt{2}\tau_j \displaystyle\sum_{l=1}^{N_c} g_{kl} \mathcal{N_F}(j,l) \nonumber \\
& & +\frac{\sigma(k)}{2} \sqrt{2}\tau_k \displaystyle\sum_{l=1}^{N_c} g_{jl} \mathcal{N_F}(k,l) \label{eqn:covFP}
\end{eqnarray}
Again, this equation is similar to parts of Eq.~\eqref{reslt_covr}. When $j=k$ in Eq.~\eqref{eqn:covFP}, we recover Eq.~\eqref{eqn:varFP}.

\begin{figure*}[tb] 
\begin{center}
	\includegraphics[width=6in]{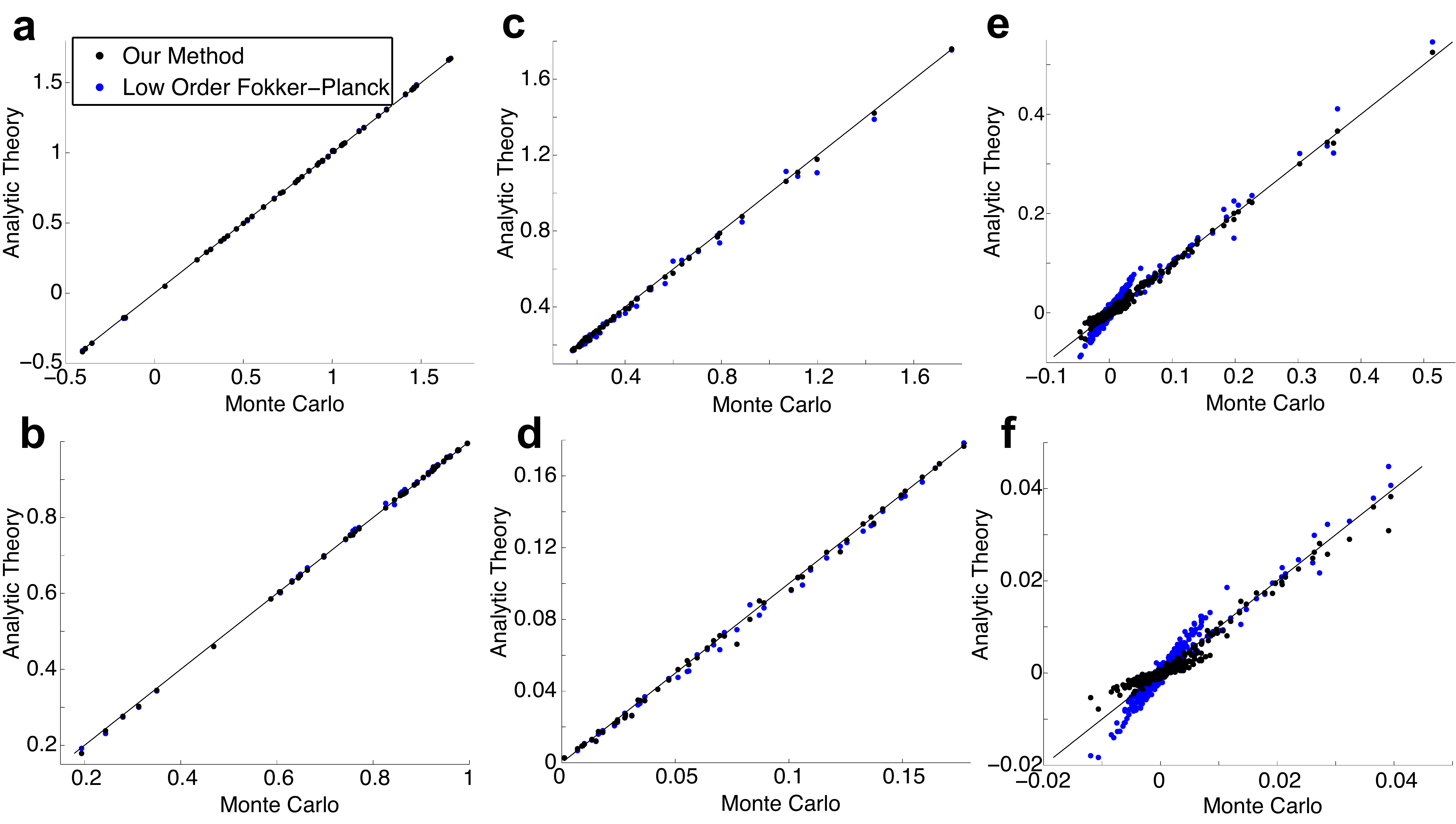}
\end{center}
\caption{Comparing our method with the lowest order approximation of the Fokker-Planck equation (Eqs~\eqref{eqn:meanFP},~\eqref{eqn:varFP},~\eqref{eqn:covFP}).  
The network configuration is described in section~\ref{sec:test_delta} 
 and is the same realization in Fig.~\ref{fig5}.  There are $N_c=50$ neurons with most parameters fixed except the time-scale $\vec{\tau}$ and coupling matrix {\bf G}.  
(a) Comparison of the mean activity $\mu(j)$ calculated via Monte Carlo simulations (horizontal axis) and our reduction method (black) and the Fokker-Planck approximation (blue), showing all 50 values for each.  
(b) Similar to (a) but for mean firing rate $\nu_j$.  
(c) Variance of activity $\sigma^2(j)$.  
(d) Variance of firing rate $Var(\nu_j)$.  
(e) Covariance of activity $Cov(j,k)$, showing all 50*49/2=1225 values for each.  
(f) Covariance of the firing rate $Cov(\nu_j,\nu_k)$.  Our result is more accurate, especially for the second order statistics (c-f).  
\label{fig7}}
\end{figure*}

Together, Eqs.~\eqref{eqn:meanFP},~\eqref{eqn:varFP}, and~\eqref{eqn:covFP} form a system of transcendental equations for the complete set of first and second order statistics.  This can be thought of as a 
{\bf lowest order approximation} to the exact statistics of the coupled system.  We implemented this method on the same network described in Section~\ref{sec:test_delta}
 and found that it is not as accurate as our method (Fig.~\ref{fig7}, black dots are closer to the diagonal line than blue dots).  
 The mean firing rates and activity perform equally well with both methods (Fig.~\ref{fig7}(a,b)), but our method outperforms this method in calculating the 
 variances (Fig.~\ref{fig7}(c,d)) and even more so with the covariances (Fig.~\ref{fig7}(e,f)).  
 
\subsection{Higher order moment closure methods}

To derive a higher order moment closure method, one can 
continue the procedure 
described in the previous section, 
by multiplying $x_i x_j x_k$ with Eq.~\eqref{fp_sseqn} and devise a method to close the lower order equations.  
Equation~\eqref{eqn:meanFP} remains the same because of the underlying normal distribution assumption:
$$\mu(j) = \mu_j + \sum_{k=1}^{N_c} g_{jk} \mathcal{E}_1(k). $$
The second set of equations (obtained by multiplying by $x_j x_k$ and integrating in $\vec{x}$) can possibly be used to better approximate higher order equations, rather than close it as was done in the {\bf lowest 
order approximation}.  

If one were to follow the outline of this method, the Gaussian assumptions on $\vec{x}$ reduce the higher moments $\int d\vec{x} x_i x_j x_k p(\vec{x})$ in terms of the mean and (co-)variances of $\vec{x}$, 
resulting in an over-constrained or redundant system.  A possible way to proceed is to devise an approximation to $\int x_j F_k(x_k)  p(\vec{x})\,d\vec{x}$, possibly relating to $\int x_j^2 F_k(x_k)  p(\vec{x})\,d\vec{x}$, where 
an assumption beyond the Gaussian approximation of $\vec{x}$ could perhaps be used.  
Whether or not there exists a higher order moment closure method on the Fokker-Planck equation~\eqref{fp_sseqn} that is more accurate than our new method is beyond the scope of this study and an interesting 
area for future research.  

What is very clear through all these calculations is that our method described in the main text is different than any common moment closure methods on the Fokker-Planck equation, 
despite some similarities in the equations.  From 
the moment closure methods we have outlined here, we see that the resulting equations will never have second order terms in the network coupling (i.e., $F_k(x_k)^2$ or $F_j F_k$), and is thus a different approach 
than our method.

%


\end{document}